\documentclass[a4paper,11pt]{article}
\usepackage{pos}
\usepackage{amsmath,bm,mathrsfs}
\usepackage{graphicx,hyperref}
\usepackage{color}

\newcommand{\imag}{\mathrm{i}}

\newcommand{\aux}{t}

\title{Scales in light-nuclei production near the QCD critical point }
\author*[a,b,c]{Shanjin Wu}
\author[d,a]{Koichi Murase}
\author[b,c,a]{Huichao Song}

\affiliation[a]{Center for High Energy Physics, Peking University, Beijing 100871, China}
\affiliation[b]{School of Physics and State Key Laboratory of Nuclear Physics and
Technology, Peking University, Beijing 100871, China}
\affiliation[c]{Collaborative Innovation Center of Quantum Matter, Beijing 100871, China}
\affiliation[d]{Yukawa Institute for Theoretical Physics, Kyoto University, Kyoto 606-8502, Japan}

\emailAdd{shanjinwu2014@pku.edu.cn}
\emailAdd{koichi.murase@yukawa.kyoto-u.ac.jp}
\emailAdd{huichaosong@pku.edu.cn}

\abstract{Based on the coalescence model, we analyse
the light-nuclei production near the critical
point by expanding the phase-space distribution function $f(\bm{r},\bm{p})$
in terms of the phase-space cumulants $\sim \langle r^m p^m\rangle_c$.
We show that the dominant contribution
of the phase-space distribution to the yield of light nuclei
is determined by the second-order phase-space cumulants.
Here, we identify the fireball size, the homogeneity length, and the effective temperature,
which are encoded in the second-order phase-space cumulants,
as the relevant scales in explaining the yield of light nuclei.
These scales are typically much larger
than the correlation length of the critical fluctuations
created in the rapid expansion of the heavy-ion systems,
so we need to eliminate this dominant contribution of the relevant scales
in order to isolate the critical contribution from the yield of light nuclei.
We find that the second-order phase-space cumulants
appeared in the yields of light-nuclei with different mass numbers
share a similar structure.
This property allows us to construct ratios of light-nuclei yields in appropriate combinations
so that the effect of the relevant scales of the light-nuclei yield cancels, which isolates the critical effects.}

\FullConference{%
  The Tenth Annual Conference on Large Hadron Collider Physics,\\
  16-20 May 2022\\
  Online
}

\begin{document}
\maketitle

\section{Introduction}
Searching for the QCD critical point is one of the most important goals for relativistic heavy-ion collisions. %One of the characteristic properties of the critical point is the divergence of correlation length, which could potentially be imprinted in the observables of the heavy-ion experiments.
Preliminary measurements of net-proton multiplicity
fluctuations at the Relativistic Heavy Ion Collider (RHIC) show a non-monotonic behavior as a function of the colliding energy~\cite{STAR:2020tga},
which qualitatively
%\KM{The word ``approximately'' hears like semi-quantitatively, but I do not seem to find qualitative predictions in Ref.~[2]}
agrees with the theoretical prediction~\cite{Stephanov:2011pb}.
Meanwhile, the realistic dynamics of the heavy-ion collision reaction is so complicated
that it is non-trivial whether the observed non-monotonic behavior is unique to the critical effect.
Thus, it is preferable to confirm the critical effects in multiple observables. Light-nuclei production is claimed to be related to the relative neutron density fluctuations~\cite{Sun:2017xrx} and its non-monotonic behavior~\cite{STAR:2022hbp} is the imprint of critical point. In this study, we consider the impact of the critical effect on another observable,
the light-nuclei yield ratios calculated from the phase-space distribution of nucleons, $f(\bm{r},\bm{p})$,
using the coalescence model~\cite{Fries:2003kq,Zhu:2015voa}.
%% the production of light nuclei is determined by the overlap of
%% the phase-space distribution $f(\bm{r},\bm{p})$ of nucleons.
In Sec.~\ref{Sec2} we demonstrate that the size of the fireball, the homogeneity length,
and the effective temperature are the relevant scales of the light-nuclei yield.
The critical fluctuations induce additional correlations in the phase-space distributions
and thus affect the yield of the light nuclei formed near the critical point.
%% In the case the light nuclei is not sufficiently close to the critical point,
The critical correlation length in rapidly expanding heavy-ion systems
is typically much smaller than the relevant scales of the light-nuclei production.
We find that the ratios of the yield in appropriate combinations can be used
to eliminate the effect of the relevant scales
and isolate the critical effect from the light-nuclei yield.
%% We analyse the scale for the light-nuclei production near
%% the critical point and discuss the role of correlation length in the light-nuclei production.

\section{Phase-space distribution in light-nuclei production}\label{Sec2}
One of the widely used methods to calculate the production of light nuclei is the coalescence model~\cite{Fries:2003kq,Zhu:2015voa, Zhao:2018lyf,Zhao:2020irc},
in which the yield is obtained as
\begin{align}\label{eq:coale-1}
  N_A=g_A \int \biggl[\prod^A_i d^3\bm{r}_i d^3\bm{p}_if_i(\bm{r}_i,\bm{p}_i)\biggr] W_A(\{\bm{r}_i,\bm{p}_i\}^A_{i=1}),
\end{align}
where the statistical factor $g_A=(2s+1)/2^A$ is given by the spin $s$ of the light nucleus.
The probability of producing light nuclei from $A$-nucleons at the phase-space positions $(\bm{r}_i,\bm{p}_i)$ ($i=1,\dots,A$)
is described by the Wigner function of the spherical harmonic oscillator
$W_A(\{\bm{r}_i,\bm{p}_i\}^A_{i=1})=8^{A-1}\exp[-\frac{1}{2}\sum^{A-1}_{i=1}\bm{Z}_i^2]$.
The Wigner function only depends on the relative distances
$\bm{Z}_i=\sqrt{\frac{i}{i+1}}(\frac{1}{i}\sum^i_{j=1}\bm{z}_j-\bm{z}_{i+1})$ ($i = 1,\dots,A-1$)
but not on the center-of-mass motion $\bm{Z}_A=A^{-1/2}\sum^A_{i=1}\bm{z}_i$,
where we redefined the phase-space variables as $\bm{z}_i=\sqrt{2}(\bm{r}_i/\sigma_A,\sigma_A\bm{p}_i)$.
This property stems from the fact that the nuclear interactions depend on the relative coordinates
between the nucleons, namely the translational invariance of the system.
The fact that the transform between the relative distances $\bm{Z}_i$ and the nucleon positions $\bm{z}_i$ is orthogonal will play an important role later.

One of the significant consequences of the Wigner function written in the Gaussian form with respect to the relative distances
%\KM{actually the fact that the Wigner function has the Gaussian form plays the most important role here.}
is that the light nuclei constituted of different numbers of nucleons $A$
share the same structure in the case of Gaussian phase-space distributions $f_i(\bm{r}_i,\bm{p}_i)$.
To see this in a systematic manner, we expand the phase-space distribution by the phase-space cumulants~\cite{Wu:2022cbh}:
%\KM{The following is not directly the characteristic function. The characteristic function is $\phi(t) = \int dz \rho(z) e^{itz}$}
\begin{align}\label{eq:dist.cumulants}
  \frac{f(\bm{z}_i)}{N_p}
    &= \rho(\bm{z}_i) = \int \frac{d^6\bm\aux_i}{(2\pi)^6} e^{-\imag\bm\aux_i\cdot \bm{z}_i}
    \exp \biggl[\sum_{\bm{\alpha}\in\mathbb{N}_0^6} \frac{\mathcal{C}_{\bm{\alpha}}}{\bm{\alpha} !}(\imag \bm\aux_i)^{\bm{\alpha}}\biggr],
\end{align}
where $N_p = N_n = \int d^6\bm{z} f(\bm{z})$ is the number of nucleons (where isospin asymmetry is neglected).
The cumulant of the phase-space variable $\mathcal{C}_{\bm{\alpha}}\equiv\langle \bm{z}^{\bm{\alpha}}\rangle_c$
is defined by the multi-index order $\bm{\alpha}\in\mathbb{N}_0^6$,
and $\langle\cdots\rangle = (1/N_p)\int d^6\bm{z} \cdots f(\bm{z})$ represents the average over the phase-space under a
single phase-space distribution $f(\bm{r},\bm{p})$. When the distribution is sufficiently close to the Gaussian
distribution, i.e., $\mathcal{C}_{\bm{\alpha}}$ for $|\bm{\alpha}|\ge3$ are sufficiently small, the yield of light
nuclei in Eq.~\eqref{eq:coale-1} can be diagrammatically evaluated by the perturbation to the Gaussian integration
and finally gives the form:
\begin{align}\label{eq:NA-factorized}
  N_A &= g_A N_p \biggl[\frac{8 N_p}{\sqrt{\det(\mathcal C_2 + \mathcal I_6)}}\biggr]^{A-1}\cdot
    [1 + \mathcal O (\{\mathcal C_{\bm\alpha}\}_{|\bm\alpha|\ge3})],
\end{align}
Here, one can see that, at the lowest order of the perturbation determined by the second-order cumulants $\mathcal C_2$,
the phase-space distribution $f(\bm{r},\bm{p})$ plays a similar role in light-nuclei yields of different mass numbers $A$.
Consequently, we may construct the ratios of the light-nuclei yields,
which are fixed solely by $g_A$ (under the assumption of the common light-nuclei size $\sigma_A\equiv \sigma$):
\begin{align}\label{ratio}
  R_{A,B}^{1-B,A-1} =
  \frac{N_p^{B-A} N_B^{A-1}}{N_A^{B-1}}
  &=\frac{g^{A-1}_B}{g_A^{B-1}}[1+\mathcal{O}(\{\mathcal{C}_{\bm{\alpha}}\}_{|\alpha| \ge 3})],
\end{align}
where the dominant contribution from the second-order phase-space cumulants is canceled out.
Explicitly, the canceled second-order cumulants have the form:
\begin{align}
  \mathcal{C}_2&=2\begin{pmatrix}
    \frac{\langle \bm{r}\bm{r}^\mathrm{T}\rangle}{\sigma^2} &
    \langle \bm{r}\bm{p}^\mathrm{T}\rangle \\
    \langle \bm{p}\bm{r}^\mathrm{T}\rangle &
    \sigma^2 \langle \bm{p}\bm{p}^\mathrm{T}\rangle
  \end{pmatrix},
\end{align}
where $\langle\bm{a}\bm{b}^\mathrm{T}\rangle$ is the $3\times3$ matrix with the elements
$\langle a_i b_j\rangle_c$ ($i,j=x,y,z$). The diagonal elements are the variances of coordinates $\langle
r^2_i\rangle_c$ and momenta $\langle p^2_i\rangle_c$, corresponding to the fireball size and the effective temperature of
the nucleon spectrum, respectively.  The non-diagonal elements are the correlation between $r$ and $p$, which is related to
the homogeneity length $l$ \cite{Scheibl:1998tk}.
To summarize, we can treat the fireball size, the homogeneity length, and the effective temperature
as the relevant scales of the light-nuclei production.

\section{Critical fluctuations in light-nuclei production}
For the light nuclei created in the vicinity of the critical point, the nucleons interact with the chiral field
$\sigma(\bm{r})$, and their masses are modified with a small deviation $\delta m=g_\sigma \sigma$ to the leading order.
Their phase-space distribution thus contains the correction term $\delta f$~\cite{Jiang:2015hri}:
\begin{align}\label{Eq_f}
    f=f_0+\delta f=f_0[1-g_\sigma \sigma/(\gamma T)],
\end{align}
where $f_0=f_\sigma|_{\sigma=0}$ denotes the background distribution without the critical contribution, and
$\gamma=\sqrt{\bm{p}^2+m^2}/m$ is the Lorentz factor. In addition to the contribution from the background $f_0$, the
polynomials of the correction term $\delta f$ also play a role in the yield of light nuclei in Eq.~\eqref{eq:coale-1}. As the
first step of the study, let us borrow the forms of the static critical correlators~\cite{Jiang:2015hri}:
\begin{align}
    &\langle \sigma(\bm{r}_1)\sigma(\bm{r}_2)\rangle_\sigma=TD(\bm{r}_1-\bm{r}_2),\\
    &\langle \sigma(\bm{r}_1)\sigma(\bm{r}_2)\sigma(\bm{r}_3)\rangle_\sigma=-2T^2\lambda_3\int d^3uD(\bm{r}_1-\bm{u})D(\bm{r}_2-\bm{u}) D(\bm{r}_3-\bm{u}),
\end{align}
where the critical propagator $D(\bm{r}_1-\bm{r}_2)\equiv \frac{1}{4\pi r}e^{-r/\xi}$
is written by $r=|\bm{r}_1-\bm{r}_2|$, the correlation length $\xi$,
and the coupling constant $\lambda_3$ for the 3-point correlator.
$\langle\cdots\rangle_\sigma$ represents the event-by-event averaging over different realizations of the sigma field.
%\KM{The result of this proceedings doesn't contain $\lambda_4$, so I've removed it.}

The interaction with the sigma field induces the critical correlation,
and the correlation length $\xi$ affects the yield of light nuclei.
This can be seen by using the characteristic function of the phase-space distribution with the critical contribution~\cite{Wu:2022sec},
where the final result takes the form:
\begin{align}\label{Eq_NA}
  \langle N_A \rangle_\sigma=g_A8^{A-1}N^A_p[\det(\mathcal{C}_2+\mathcal{I}_6)]^{-(A-1)/2}[1+\tilde{\Xi}(A)].
\end{align}
Here we defined $\tilde{\Xi}(A)\equiv\sum_{b=2}^A(-1)^bC^b_A\Xi(A,b)$, where $C^b_A$ is the binomial coefficients
and $\Xi(A,b)\sim g^b_\sigma \langle\prod^b_{j=1}\sigma(\bm{t}_{r,j})\rangle_\sigma$ is the critical contribution
which encodes the critical correlation length $\xi$.
Although the correlation length grows up to $\xi = 1/m_\sigma$ in the static systems with $m_\sigma$ being the $\sigma$ mass,
the correlation length is limited to the order of 2--3 fm
%\KM{$\sim$ is only used for the range in CJK}
in heavy-ion collisions due to the rapid expansion of the system~\cite{Berdnikov:1999ph},
which is typically much smaller than the relevant scales of the light-nuclei production
encoded in $\mathcal{C}_2$. Considering the small value of $\xi$ and small critical regime on the QCD phase diagram, the
correlation length in the individual light-nuclei yield as shown in Eq.~\eqref{Eq_NA} is negligible. However, due to
the similar structure related to the second-order phase-space cumulants $\mathcal{C}_2$ in Eq.~\eqref{Eq_NA}, the combination such as
\begin{align}\label{Eq_RAB}
  \tilde{R}(A,B)&\equiv R^{1-B,A-1}_{A,B}-g^{A-1}_Bg^{-(B-1)}_A
\end{align}
and
\begin{align}\label{Eq_RABD}
    \tilde{R}(A,B,D)\equiv R^{1-B,A-1}_{A,B}-g^{A-1}_Bg^{-(A-1)(B-1)/(D-1)}_D[R^{1-D,A-1}_{A,D}]^{(B-1)/(D-1)}
\end{align}
greatly suppress the contribution from the background scales in $\mathcal{C}_2$ which help to isolate the effects related to
the correlation length. Here, the definition of $R^{1-B,A-1}_{A,B}$ in Eq.~\eqref{ratio} is adapted to the present case as
$R^{1-B,A-1}_{A,B}\equiv\langle N_B\rangle_\sigma^{A-1} \langle N_A\rangle^{-(B-1)}_\sigma N_p^{B-A}$.

\section{Summary}
In this study, we discussed the light-nuclei production near the QCD critical point
from the viewpoint of the relevant scales of the light-nuclei production $\mathcal C_2$ and the scale of the critical correlation length $\xi$.
We first decomposed the phase-space distribution $f(\bm{r},\bm{p})$ in terms of various orders of phase-space cumulants $\mathcal C_{\bm{\alpha}}$.
Since the Wigner function in the coalescence model is approximately written in Gaussian form
with respect to the relative phase-space distances of constitutive nucleons,
the yield of light nuclei share a similar structure at the lowest order of the phase-space cumulants,
We identified the relevant scales of the yield in the second-order cumulants: the fireball size, the homogeneity length, and the effective freeze-out temperature.
The phase-space distribution of nucleons is modified by the interaction with the chiral field,
which would be reflected in the yield of light nuclei.
Naively, it would seem hard to separate the critical contributions from the relevant scales of the background phase-space distribution,
but the structure of the lowest order of the phase-space cumulants in the yield
enables us to construct combinations of light-nuclei yields
to suppress the relevant scales encoded in the second-order phase-space cumulants and
isolate the effects of the correlation length.

\end{document}